\documentclass[aps,prl,twocolumn,unsorteddaddress,longbibliography,showpacs,amsmath,amsmath,amssymb,amsfonts,notitlepage,superscriptaddress]{revtex4-1}

\usepackage[utf8]{inputenc}

\usepackage{xcolor}
\usepackage{amsmath}
\usepackage{amsfonts}
\usepackage{amssymb}
\usepackage{makeidx}
\usepackage{graphicx}
\usepackage{tikz}
\usepackage{float}
\usepackage{braket}
\usepackage{lipsum}
\usepackage{enumitem}
\usepackage{comment}

\usepackage[caption=false]{subfig}

\usepackage{color}
\usepackage[colorlinks=true,citecolor=blue,linkcolor=blue,urlcolor=blue]{hyperref}
\usepackage{url}

\usepackage{braket}
\usepackage{verbatim}

\DeclareMathOperator{\tr}{\text{Tr}}

\newcommand{\op}[2]{\ket{#1}\!\bra{#2}}
\newcommand{\beq}{\begin{equation}} 							
\newcommand{\eeq}{\end{equation}}
\newcommand{\bematrix}{\left(\begin{matrix}}
\newcommand{\ematrix}{\end{matrix}\right)}

\begin{document}

\title{Optimizing quantum transport via the quantum Doob transform}

\author{Dolores Esteve}
\email[]{desteve@onsager.ugr.es}
\affiliation{Quantum Thermodynamics and Computation Group, Departamento de Electromagnetismo y F\'isica de la Materia, Universidad de Granada, Granada 18071, Spain}

\author{Carlos P\'erez-Espigares}
\email[]{carlosperez@ugr.es}
\affiliation{Departamento de Electromagnetismo y F\'isica de la Materia, Universidad de Granada, Granada 18071, Spain}
\affiliation{Institute Carlos I for Theoretical and Computational Physics, Universidad de Granada, Granada 18071, Spain}

\author{ Ricardo Guti\'errez}
\email[]{rigutier@math.uc3m.es}
\affiliation{Universidad Carlos III de Madrid, Departamento de Matem\'aticas, Grupo Interdisciplinar de Sistemas Complejos (GISC), Avenida de la Universidad, 30 (edificio Sabatini), 28911 Leganés (Madrid), Spain}

\author{Daniel Manzano}
\email[]{manzano@onsager.ugr.es}
\affiliation{Quantum Thermodynamics and Computation Group, Departamento de Electromagnetismo y F\'isica de la Materia, Universidad de Granada, Granada 18071, Spain}
\affiliation{Institute Carlos I for Theoretical and Computational Physics, Universidad de Granada, Granada 18071, Spain}

\begin{abstract}
Quantum transport plays a central role in both fundamental physics and the development of quantum technologies. While significant progress has been made in understanding transport phenomena in quantum systems, methods for optimizing transport properties remain limited, particularly in complex quantum networks. Building on recent advances in classical network optimization via the generalized Doob transform, we introduce a novel method that extends this approach to quantum networks. Our framework leverages a single diagonalization of the system generator to efficiently tailor both the Hamiltonian and dissipative contributions, optimizing transport observables such as currents and activities. We demonstrate the method’s effectiveness through extensive numerical explorations, showing that optimal performance arises from non-trivial modifications to both coherent and incoherent dynamics. We also assess the robustness of the optimization under constraints that preserve specific physical features, such as fixed dissipative structures and input-output interactions. Finally, we discuss the connection between  optimized transport and centrosymmetry, highlighting the relevance of this property for enhanced transport efficiency in quantum systems.
\end{abstract}
\maketitle

{\bf{Introduction:}} Quantum transport phenomena has been an important field of research since the beginning of quantum theory itself, a field that has grown in recent decades due to its importance for the development of quantum devices \cite{dubi:rmp11}. From a theoretical perspective, the study of quantum transport has helped understanding Fourier's law in both linear \cite{manzano:pre12,znidaric:pre11} and lattice \cite{asadian:pre13,znidaric:prl13,manzano:njp16} systems. To date, one finds an extensive study of transport phenomena in quantum networks \cite{scholak:pre11,mohseni:jcp08,chin:njp10,manzano:po13,scholak:jpb11, cao:jpc09,moix:njp13,Walschaers:prl13}.

Recently, a method to optimize transport in classical networks has been proposed \cite{gutierrez21a, gutierrez21b}. It is based on the generalized Doob transform \cite{jack2010,chetrite2015} of random walks on weighted graphs \cite{masuda2017,riascos21}, which relies on the analysis of large deviations in such systems \cite{debacco2016,coghi2019}. Given an adjacency matrix describing all possible transitions, it allows for tailoring the transition rates in order to optimize transport observables including activities or currents \cite{gutierrez21b}, as well as to find generalized optimal paths defined in terms of the statistics of such dynamical observables \cite{gutierrez21a}.

In this paper, we introduce a method to optimize transport properties of quantum networks based on the quantum Doob transform, which extends that theoretical framework for transport optimization on classical networks to the quantum realm. Our method can obtain better performing networks with a single diagonalization of the system generator, improving dramatically the computational cost in comparison with methods previously proposed. These methods are based on Monte Carlo sampling or genetic algorithms and require the analysis of a high number of systems to find an optimal one. 

In contrast, our method relies on a small number of transformations of a system based on the above-mentioned diagonalization.  This allows us to design a new Lindblad dynamics by redefining the Hamiltonian and jump operators. This new Liouvillian presents as typical the behavior that was exotic in the original one, allowing us to increase the efficiency of the network. By studying its performance, we show that the optimized systems undergo non-trivial changes in both the Hamiltonian and dissipative parts. We also test the method under certain constrains, as maintaining the dissipative part and the input-output interaction constant, to validate its robustness. Besides, we investigate the centrosymmetry, a property connected to transport efficiency \cite{Walschaers:prl13,zech:njp14}, of the optimized networks. This work presents, to our knowledge, the first application of the quantum Doob transform to optimize quantum transport, and, more generally, to complex systems beyond qubits and qutrits. It also opens the door to new large-deviation techniques for optimizing multiple observables, finding optimal paths, and unveiling dynamical phase transitions in transport phenomena, such as those recently proposed for classical networks \cite {gutierrez21a,gutierrez21b}.

{\bf Methodology:} The density matrix $\rho(t)$ of a Markovian quantum system, weakly coupled to an environment, is described by the Gorini–Kossakowski–Sudarshan–Lindblad (GKSL) quantum master equation $\dot{\rho}(t)=\mathcal{L}[\rho(t)]$ \cite{breuer2002,gardiner2004,lindblad1976,gorini1976,benatti2005,alicki2007,baumgartner2008,manzano2020,landi_rmp_22}, where 

\begin{equation} \label{eqn:master}
\begin{split}
\mathcal{L}[\cdot]&\equiv-i [H,\cdot]+\sum_{i=1}^{d}\left(L_i \cdot L^\dagger_i-\frac{1}{2}\left\{L^\dagger_i L_i,\cdot\right\}\right)
\end{split}
\end{equation} 
is the Liouvillian superoperator $\mathcal{L}$ (with $\hbar=1$), which includes the Hamiltonian $H$ that describes the system's coherent dynamics, and $d$ jump operators $L_i$ encoding the dissipative effects due to interactions with the environment. 
\begin{figure}[h]
\includegraphics[scale=0.18]{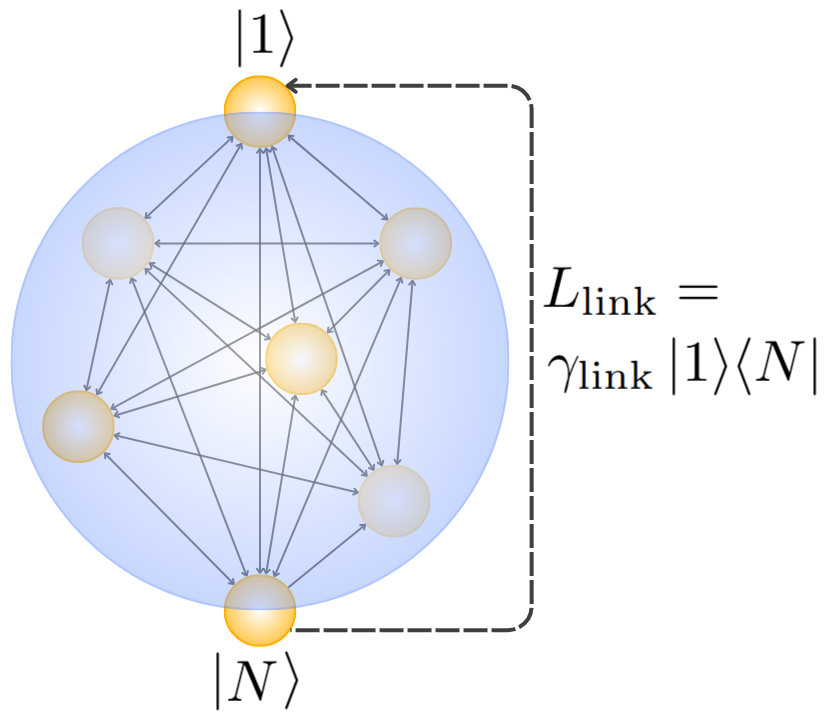}
\caption{Sketch of a fully connected quantum network with an incoherent link between two nodes.}
\label{fig:network}
\end{figure}

Our system of interest is a fully connected network of two-level systems as displayed in Fig \ref{fig:network}. Quantum networks are a mainstream field of study with many applications beyond the reach of classical ones \cite{Meter_14,Azuma_23}. This specific kind of quantum networks have been widely used to study different problems such as energy transport in photosynthetic complexes \cite{scholak:pre11,mohseni:jcp08,chin:njp10,manzano:po13,scholak:jpb11}, computational science analysis \cite{sanchez-burillo:sr12,Paparo2012}, and optimization of transport phenomena \cite{cao:jpc09,moix:njp13,Walschaers:prl13}. We work in the single-excitation manifold, described by the Fock basis $\left\{ \left|i\right> \right\}_{i=1}^N$. In this basis the Hamiltonian can be written as 

\begin{equation}
H= \sum\limits_{i<j} J_{ij}\op{j}{i} + \text{H.c.},
\end{equation}
where the elements $J_{ij}$ represent the coupling strengths between the sites. This form of Hamiltonian is totally general beyond the case studied here as it correspond to any $N$ level system. This means that the results presented in this paper can be extended to arbitrary systems with both coherent and incoherent transitions.

In our model, inspired by Refs. \cite{scholak:pre11,manzano:po13}, two of the sites ($\ket{1}$ and $\ket{N}$) are taken as the input and output sites, meaning that we focus on excitations  as they travel from $\ket{1}$ to $\ket{N}$. To study the efficiency of transport in this setting we include a single incoherent link in the form $L_\textrm{link}=\gamma_\textrm{link}\op{1}{N}$, the flux of excitations through it being an indication of energy transfer across the network; in Eq.~(\ref{eqn:master}), $d=1$ and $L_1 = L_\textrm{link}$. In a more general framework there could be several incoherent channels in the system, and we can apply the Doob transform to an arbitrary number of them. Even if the technique is the same, the results may be complicated, as the description of coupled incoherent observables presents very specific features  \cite{manzano21}.

In Refs. \cite{Walschaers:prl13,zech:njp14} the energy transfer across a disordered network is related to the amount of {\it centrosymmetry} in the structure of the Hamiltonian with respect to $\ket{1}$ and $\ket{N}$. More efficient systems are  characterized by invariance under permutation of the intermediate sites. Specifically, centrosymmetry is defined as 

\begin{equation}
\varepsilon = \frac{1}{N} \min_{\cal{S}} \left\| H - A^{-1} H A \right\|,
\label{eq:centrosym}
\end{equation}
where $\left\| O \right\|= \sqrt{\text{Tr}\left[ O^\dagger O \right]}$ is the Hilbert-Schmidt norm of an operator and $A$ is the exchange matrix $A_{i,j} = \delta_{i, N-j+1}$ that exchanges site $N$ with $1$, $2$ with $N-1$, and so on. Finally, the measure is minimized over all possible permutations $\cal{S}$ of the intermediate sites $2,\,...,\,N-1$. Below we will relate the centrosymmetry of network Hamiltonians to the optimization of transport resulting from the large-deviation methodology we describe next.

 Quantum transport will be optimized by the use of a technique known as the generalized Doob transform applied to open quantum systems \cite{Carollo2018}. This technique modifies the transition rates (both coherent and incoherent) in a given dissipative quantum system in order to make rare events typical, which we will exploit to optimize transport on the quantum network described above, much as the classical Doob transform allows for the optimization of classical networks \cite{gutierrez21a}.  We consider a scalar observable $\mathcal{O}$ given by the number of incoherent events along a trajectory, each one corresponding to one transition $\ket{N} \to \ket{1}$ associated with the operator $L_\textrm{link}$. The goal is to derive a new master equation, also in GKSL form \eqref{eqn:master}, whose stationary dynamics naturally display as typical enhanced transport efficiencies given by large average values $\braket{\mathcal{O}}$, which are rare (their probability being exponentially suppressed with time) in the original dynamics. 

Conceptually, this is achieved by modifying the exponentially decaying probability distribution of $\mathcal{O}$ in a trajectory of duration $t$, denoted as $P_t(\mathcal{O})\approx e^{-t I(\mathcal{O}/t)}$ [here $I(\mathcal{O}/t)$ is the large-deviation function, which encodes the statistics of rare events \cite{Touchette2009}], so that it becomes exponentially biased (or ``tilted'') by means of a conjugate parameter $s$, $P_t^s(\mathcal{O}) = e^{s  \mathcal{O}} P_t(\mathcal{O})/Z_t(s)$. The normalizing factor $Z_t(s)$ is nothing but the moment generating function of $P_t(\mathcal{O})$, and for long times reads $Z_t(s)\approx e^{t \theta(s)}$, with $\theta(s)$ being the scaled cumulant-generating function (SCGF). The tilted distribution, $P_t^s(\mathcal{O})$, which is generated by an ensemble of trajectories known as the $s$-ensemble \cite{garrahan2009}, yields values of $\mathcal{O}$ larger (for $s>0$) or smaller (for $s<0$) than typical ($s=0$) in the original dynamics, as the case may be. 

As our interest lies mainly in enhancing transport efficiency, we will mostly focus on $s>0$. Such biased distribution is exponentially hard to sample, yet its SCGF $\theta(s)$, which contains the full statistics of $\mathcal{O}$ for all $s$, can be obtained from the so-called tilted Liouvillian  \cite{garrahan10a}, 

\begin{equation}
\resizebox{\columnwidth}{!}{$\mathcal{L}_s[\cdot] = - i [H,\cdot] + \left(e^{s} L_\textrm{link} \cdot L_\textrm{link}^\dag -\frac{1}{2} \left\{L_\textrm{link}^\dag L_\textrm{link},\cdot \right\}\right),$} \label{eqn:tilted}
\end{equation}
differing from the original dynamics in the $e^{s}$ factor appearing in the first term of the dissipative part. In fact, $\theta(s)$ corresponds to the eigenvalue of $\mathcal{L}_s$ \eqref{eqn:tilted} with the largest real part, and it is a real-valued function that is numerically obtained for different values of $s$ \cite{garrahan10a,manzano:av18}.

While Eq.~\eqref{eqn:tilted} does not correspond to a physical evolution as it is  not trace-preserving, its spectral properties allow us to choose a tilting parameter value $s$ yielding the statistics of interest through the derivatives of $\theta(s)$ [i.e.~the cumulants of $P_t^{s}(\mathcal{O})$]. For our purposes, the most important is the first scaled cumulant or current $J(s) = \braket{\mathcal{O}}_{s}/t$, where the brackets denote averaging with respect to $P_t^{s}(\mathcal{O})$, which is given by the first derivative of the SCGF, $J(s) = \theta'(s)$.  The diagonalization of Eq.~\eqref{eqn:tilted} moreover allows us to construct the quantum Doob transform, i.e.\! a physical dynamics that naturally produces such statistics corresponding to $s\neq 0$ in its stationary state. In this regard, the left eigenvalue problem $\mathcal{L}^\dagger_{s}[l_{s}] = \theta(s)\, l_{s}$, where $l_{s}$ is the (left) eigenmatrix associated with the eigenvalue with the largest real part of the adjoint tilted operator $\mathcal{L}^\dagger_{s}$, is particularly important  (see SM for details). Indeed, every term in the resulting quantum master equation, whose stationary-state statistics for $\mathcal{O}$ is distributed following $P_t^{s}(\mathcal{O})$, can be obtained from the original Hamiltonian $H$ and jump operator $L_\textrm{link}$ by a suitable transformation involving $l_{s}$. Hence, the effective (Doob) Hamiltonian reads

\begin{equation}
H^D_{s} = \frac{1}{2}\, l_{s}^{1/2}\left(H- \frac{i}{2}  L^{\dag}_\textrm{link} L_\textrm{link} \right) l_{s}^{-1/2} + \text{H.c.},
\label{eq:hd}
\end{equation}
and the effective (Doob) jump operator is given by 
\begin{equation}
 L^D_{s} = e^{s/2 }\, l_{s}^{1/2}\, L_\textrm{link}\, l_{s}^{-1/2}.
 \label{eq:ld}
\end{equation}
This may now include linear combinations of several incoherent transitions, including some absent from the original operator $L_\textrm{link}$. A more detailed discussion of this framework, with possibly several incoherent transitions in the network and for general scalar trajectory operators, is provided in the Supplemental Material (SM). Notice that for $s = 0$, we recover the original dynamics, with $r_{s=0}$ being the stationary state, satisfying $\mathcal{L}_{s=0}[r_{s=0}] = 0$ (since $\theta(0) = 0$). The corresponding left eigenmatrix is $l_{s=0} = \mathcal{I}$, as $\mathcal{L}_{s=0}^{\dagger}[\mathcal{I}] = 0$, which reflects the conservation of probability under $\mathcal{L}_{s=0}$.

\begin{figure}[h]
    \includegraphics[width=1\linewidth]{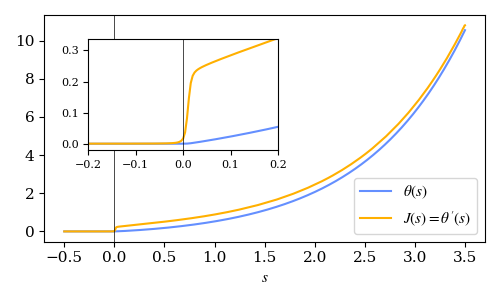}
    \includegraphics[width=1\linewidth]{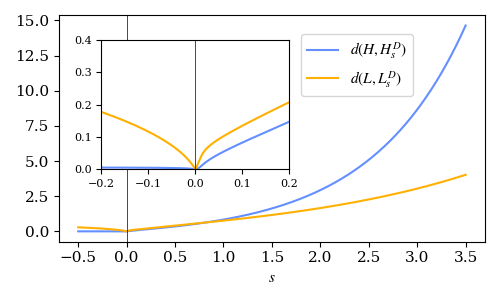}
    \caption{Top: Scaled cumulant-generating function $\theta(s)$ and current $J(s) = \theta'(s)$, as a function of the tilting parameter $s$. Bottom: Trace distance between the Doob Hamiltonian \eqref{eq:hd} and Doob jump operator \eqref{eq:ld} and their corresponding original ($s=0$) operators. Both plots are based on the system with highest improvement over $M = 10^4$ randomly generated Hamiltonians.}
    \label{fig:Doob_Networks}    
\end{figure}

{\bf Results:} To put this framework into practice in a transport optimization task, we have generated $M=10^4$ uniformly distributed random  Hamiltonians $H$ of $N=7$ sites \footnote{The value $N=7$ has been selected as it is very popular in quantum transport, since it coincides with the number of chromophores of the Fenna-Mathew-Olson complex. We have checked that the same conclusions are valid for larger systems up to $N=10$ nodes.}, and studied its transport properties under the Doob transform (see SM for details about the random generation). First, we have selected the configuration with the highest improvement under a Doob transform with $s=3.5$ \footnote{The choice $s=3.5$ allows us to study a very substantial improvement in transport efficiency without yet encountering numerical difficulties. The latter arise in the diagonalization of the tilted Liouvillian \eqref{eqn:tilted}, which becomes computationally challenging (due to the high numbers involved and the possibility of overflow) for larger $s$.}. For this configuration, in Fig.~\ref{fig:Doob_Networks} (top) the current $J(s)$ is displayed, together with the SCGF $\theta(s)$, both of them represented as functions of $s$ for convenience. We can observe that both present a monotonic increase with $s$, with a remarkable dynamical crossover displayed by the current at $s=0$, meaning that transport can be substantially optimized for $s>0$. In the lower panel, we display the amount of change of the Hamiltonian and jump operators measured by the trace distance \cite{nielsen_00}. Interestingly, for small values of $s$ both the dissipative and the coherent generators change in a similar way, but around $s=1$ this trend changes and the Hamiltonian plays a dominant role. 

\begin{figure}
    \centering
    \includegraphics[width=1\linewidth]{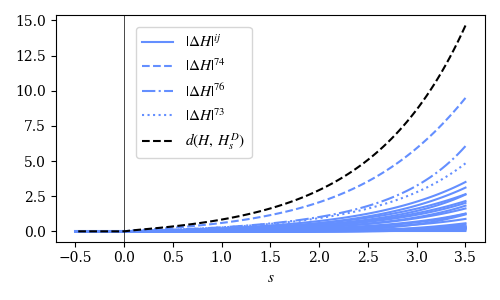}
    \includegraphics[width=1\linewidth]{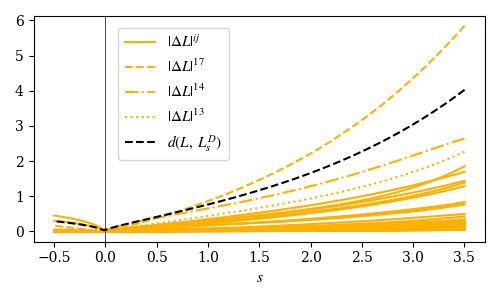}
    \caption{Differences (in absolute value) between the entries of Doob Hamiltonian $H_s^D$ \eqref{eq:hd} (upper panel) or the Doob jump operator $L_s^D$ \eqref{eq:ld} (lower panel) and those of the corresponding original operators ($s=0$), together with the trace distance between the full operators in each case.} 
    \label{fig:deviations}
\end{figure}

An important goal is to understand more deeply what kind of changes improve the efficiency of the system, and if this improvement is robust under physical constrains. To this end, in Fig.~\ref{fig:deviations} we present the change undergone by different entries of the Hamiltonian (upper panel) and the jump operator (lower panel). For this specific network, the larger modifications are in the interaction between nodes 4 and 7, for the Hamiltonian part, and between nodes 1 and 7 for the incoherent dynamics. Similar features arise for different randomly generated systems, with hardly any modification in the Hamiltonian connection between input and output (not shown). This highlights the non-trivial transformation of the dynamics performed by the Doob transform, as the Hamiltonian modifications are dominant for large $s$, see Fig.~\ref{fig:Doob_Networks} (bottom). 

\noindent

\begin{figure*}
    \centering
    \includegraphics[width=.32\linewidth]{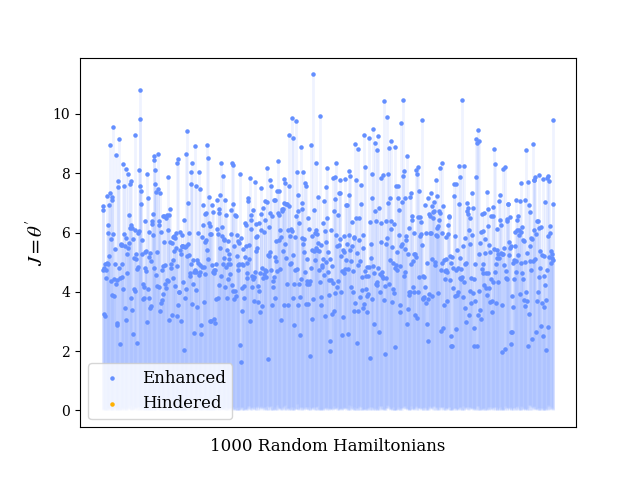}    
    \includegraphics[width=.32\linewidth]{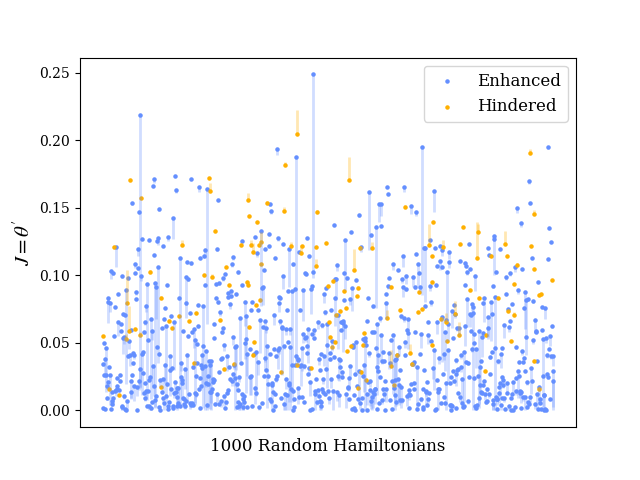}
    \includegraphics[width=.32\linewidth]{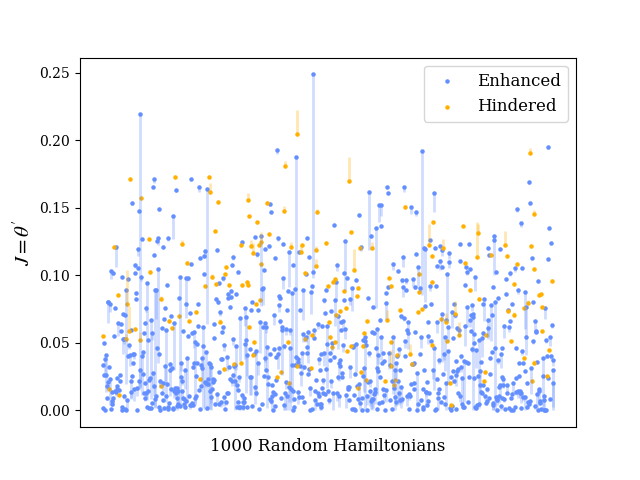}
    \caption{In all plots vertical segments start at the original effiency and end at the Doob-transformed efficiency, highlighted by a dot, with colors indicating whether the efficiency is enhanced or hindered (see legend). Left: Original and Doob-transformed efficiency, using both the Doob Hamiltonian \eqref{eq:hd} and the Doob jump operator \eqref{eq:ld}.  Center: Original and the Doob Hamiltonian together with the original jump operator $L_\textrm{link}$. Right: Original and, systems with $L_\textrm{link}$ as jump operator and the Doob Hamiltonian modified so that $(H_s^D)_{1N} = 1$. All plots are based on 1000 random initial Hamiltonians. }
    \label{fig:rain}
\end{figure*}

We next extend our analysis to the full set of $M=10^4$ randomly generated Hamiltonians, to each of which we apply the Doob transform with $s=3.5$.  In Fig.~\ref{fig:rain} (left), the  initial current $J$ corresponding to the original Hamiltonian is plotted together with the current of the optimized (Doob-transformed) network for 1000 realizations (more points are not displayed for clarity, but the conclusions stand). For this case, $100\%$ of the modified systems improve their efficiency in comparison with the original ones. 

Noting that the Doob transform changes both the Hamiltonian and the Lindblad operators, we can also consider physical restrictions to our method. Along these lines, we next consider the case where the incoherent part of the dynamics is not changed, to study the effect of the Doob Hamiltonian ~\eqref{eq:hd} alone on transport efficiency. The results are displayed in Fig.~\ref{fig:rain} (center). For this case, we get an improvement for $85.81 \%$ of the initial configurations. 

Finally, we can also consider the situation in which the Hamiltonian connection between the input and output nodes remains constant together with the original incoherent dynamics. For this case, the improvement happens in $82.46\%$ of the cases, see Fig.~\ref{fig:rain} (right). By an extensive Monte Carlo analysis (see SM for details), we conclude that the systems with initial lower transport efficiency are the ones more prone to be improved, while the initial configurations that already sustain large currents are those whose transport efficiency is more likely to remain unchanged or be hindered.

\noindent

\begin{figure}
    \centering
    \includegraphics[width=1\linewidth]{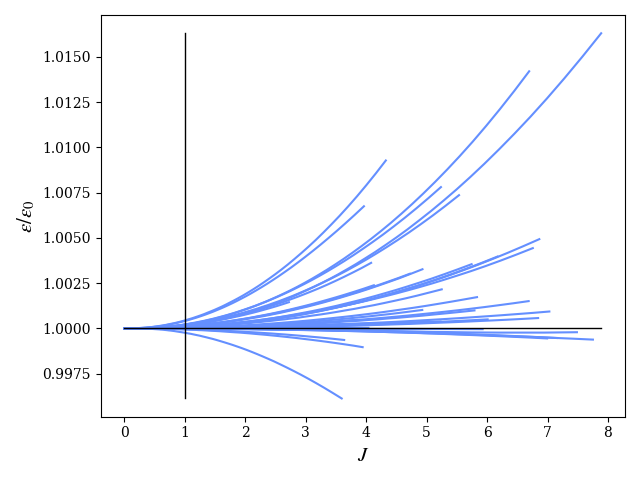}
    \caption{Centrosymmetry $\varepsilon$ \eqref{eq:centrosym} of the Doob Hamiltonians \eqref{eq:hd} normalized with respect to the centrosymmetry of the corresponding original systems ($s=0$) as a function of $J$.}
    \label{fig:centrosymmetry}
\end{figure}

According to the conclusions drawn in Refs.~\cite{Walschaers:prl13,zech:njp14} from results obtained with a very different methodology, the modifications of the Hamiltonian that enhance the efficiency should also increase the centrosymmetry of the system. To check this assumption, in Fig.~\ref{fig:centrosymmetry} we plot the  centrosymmetry $\varepsilon$ \eqref{eq:centrosym} (normalized with respect to its original, $s=0$, value) of 40 Doob-modified Hamiltonians \eqref{eq:hd} as a function of the current $J$. When $J$ increases the Hamiltonian centrosymmetry increases in most cases. We also conclude that those that are originally less centrosymmetric are more prone to increase its centrosymmetry in the Doob-modified network, indicating a strong correlation between efficiency and centrosymmetry. This is observed in the analysis of all $M=10^4$ random Hamiltonians (see SM). 

We have also estimated the computational cost of our method. For a system whose states belong to a Hilbert space of dimension $N$, a Liouvillian superoperator, see Eq.~\eqref{eqn:master}, is a (complex) matrix of dimension $N^2\times N^2$. Doob transform is based on calculating the eigenvalue with the largest real part of the tilted Liouvillian \eqref{eqn:tilted}. To do so, in the worst case scenario, we can use the Singular-Value Decomposition that has a complexity of $O(N^6)$ \cite{li:arxiv19,golub:nm70}. (In many cases, more efficient diagonalization methods can be used, as the Liouvillian may be a sparse matrix and we are interested only in the highest eigenvalue and its associated eigenvectors.) In contrast, the transport optimization in Ref. \cite{zech:njp14} is achieved through an evolutionary algorithm that generates $10^2$ Hamiltonians per step and is executed across $10^4$ steps. As the efficiency needs to be calculated for each step and Hamiltonian, this means that this approach has a complexity of $O(10^6 \,N^6)$. Moreover, in this case the full spectrum and all eigenvectors need to be computed. This comparison highlights the usefulness of our method, as it can optimize networks  through a single diagonalization, or at most the same number of diagonalizations as values of $s$ are considered. 

To conclude we touch upon two important points: the generalization of our transport optimization method to higher number of particles and its experimental viability. Transport properties strongly depend on the number of particles, the most dramatic change being that between the single-excitation manifold and the multiparticle case \cite{znidaric:prl13,manzano:njp16}. To check if our results are also applicable to the multi-excitation case, we have repeated the analysis for a two excitations system (see SM). We conclude that the optimization is very similar in that case, suggesting that our approach can be applied to more general multi-excitation networks. 

 Concerning the experimental viability of our method, in principle the Doob-modified Hamiltonian part can be implemented in different kinds of physical systems. Very recently, it has been proved that all local Hamiltonians can be simulated by the use of several universal families of Hamiltonians \cite{Zhou26}. Besides, general Hamiltonians can be simulated in quantum computers by the Trotter-Suzuki decomposition even in noisy regimes \cite{Avtandilyan25}. A more challenging task is to simulate the Doob jump operators, but it can be addressed via techniques already used to generate entangled states or invariant subspaces  \cite{kraus2008,manzano2015}. This task can also be addressed by quantum computers via dilations \cite{hu2020}.

{ \bf Conclusions:} In this paper, we have proposed a novel technique to optimize transport in quantum networks, achieving a dramatic improvement in computational complexity with respect to previous methods based on genetic algorithms. Our method is based on the Doob transform, which converts rare behavior into typical dynamics. By this approach, which modifies both the Hamiltonian and dissipative parts of the dynamics, we have shown that significant enhancements arise from non-trivial changes in the system Hamiltonian. We have also tested our method under additional constraints, such as reverting the incoherent dynamics to the original form or keeping the interaction between the input and output nodes constant. Our results demonstrate that the method is both robust and comprehensive. Finally, we have studied the role of centrosymmetry in transport optimality.

{ \bf Acknowledgements:} We acknowledge funding from the Ministry of Science and Innovation, the Ministry of Universities, the Ministry for Digital Transformation and of Civil Service, and AEI (10.13039/501100011033) of the Spanish Government through project PID2021-128970OA-I00, PID2021-123969NB-I00, PID2023-149365NB-I00, PID2024-159024NB-C21, C-EXP-251-UGR23, and QUANTUM ENIA project call - Quantum Spain project, and by the European Union through the Recovery, Transformation and Resilience Plan - NextGenerationEU within the framework of the Digital Spain 2026 Agenda, and also the FEDER/Junta de Andaluc\'{\i}a program A.FQM.752.UGR20. We are also grateful for the the computing resources and related technical support provided by PROTEUS, the supercomputing center of Institute Carlos I for Theoretical and Computational Physics in Granada, Spain.

\section*{Appendix A. Theoretical framework}

\subsection*{A.1. Dissipative quantum walks}

We consider dissipative quantum walks over graphs of $N$ vertices. Mathematically speaking, the relevant Hilbert space is $\mathbb{C}^N$ with the standard inner product, which is spanned by the canonical orthonormal basis $\{|1\rangle,|2\rangle,\ldots |N\rangle\}$, with $|j\rangle$ corresponding to vertex $j$  ($j=1,2,\ldots, N$). The state operator of such a dissipative quantum walk, $\rho(t)$ $\in \mathbb{C}^{N\times N}$, is a positive-semidefinite unit-trace Hermitian matrix, which evolves in time according to the Markovian GKSL quantum master equation, see Eq. (1) in the main text. The Hamiltonian in this basis has zero diagonal terms, while its off-diagonal elements \! $\langle k | H| j\rangle = \langle k | H| j\rangle^*$ (where $j\neq k$ and $^*$ denotes complex conjugation) indicate coherent transitions between $j$ and $k$. The jump operators can be written as linear combinations of incoherent transitions $L_i =  \sum\limits_{j} \sum\limits_{k} c^i_{jk} |k\rangle\langle j|$, with sums running over incoherently connected vertices and coefficients $c_{jk}^i \in \mathbb{C}$. 

The description given so far is valid both for the initial quantum-walk model and for the model that results from conditioning on the statistics of a given observable and applying the quantum Doob transform (see the following sections). In the initial quantum walk, the Hamiltonian $H$ is expected to be a real matrix proportional to the weighted adjacency matrix of the graph $W$, with a proportionality constant that we can take to be $1$ (any other choice would amount to a redefinition of the time units). Here $W$ is a symmetric matrix (the graph is assumed to be undirected), where $\langle k|W|j\rangle = \langle j|W|k\rangle > 0$ if $j$ and $k$ are (coherently) connected, its specific value giving the strength of the connection $|j\rangle \leftrightarrow |k\rangle$, or $0$ otherwise. Additionally, the initial jump operators are chosen to correspond to particular unidirectional transitions $|j\rangle \to |k\rangle$. We denote the corresponding jump operators $L_{j k} =  \sqrt{\gamma_{jk}} |k\rangle\langle j|$, where $\gamma_{j k}$ are the (incoherent) transition rates. 

These choices are particularly convenient for analyzing the statistics of trajectory observables of dissipative quantum walks. In fact, they constitute a generalization of the setting discussed in the main text, see Fig. 1, where the Hamiltonian is all-to-all connected (i.e.\! the matrix $W$ only has zeros along its main diagonal), with entries that are randomly assigned initially (see a more detailed description in Appendix B). Moreover, in the initial system the only jump operator $L_\textrm{link} = L_{N1}$ contains a single transition $\ket{N}\to \ket{1}$, though other possible incoherent transitions are known to emerge in the Doob-modified dissipative part \cite{Carollo2018}. The expressions that follow thus generalize those that appear in the main text, which are particularizations for $\gamma_{N1} = \gamma_\text{link}$ and $\gamma_{j k} = 0$ for any other incoherent transition (with $j\neq N$ or $k\neq 1$).

\subsection*{A.2. Time-integrated observables, tilted Liouvillian}

We consider a scalar observable $\mathcal{O}$, which counts the number of detected events along a trajectory. In our framework, an event corresponds to one or several incoherent transitions $|j\rangle \to |k\rangle$, i.e.\! the effect of one or several jump operators $L_{j k}$. The local contribution to the time-integrated observable $\mathcal{O}$ by that transition is denoted as $O_{jk}\in \{0,1\}$. Then $\mathcal{O}$, being the sum of $O_{jk}$ over all transitions in a given trajectory, is a fluctuating time-extensive observable. Its probability distribution in a trajectory of length $t$ we denote as $P_t(\mathcal{O}) = \tr\left[\rho^\mathcal{O}(t)\right]$, where $\rho^\mathcal{O}(t)$ is the density operator conditioned on having $\mathcal{O}$ events detected up to time $t$. For sufficiently long times, this distribution is expected to adopt a large-deviation form  $P_t(\mathcal{O}) \approx e^{-t I(\mathcal{O}/t)}$, with a rate function $I(\mathcal{O}/t)$ that is minimized at $\langle \mathcal{O} \rangle/t$, with $\langle \mathcal{O} \rangle = \int \mathcal{O}\, P_t(\mathcal{O})\, d\mathcal{O}$, so that values of $\mathcal{O}$ away from the average become exponentially suppressed in time.\\

By biasing or tilting these probabilities with a conjugate parameter $s$, we obtain
\begin{equation}
P_t^s(\mathcal{O}) = \frac{e^{s \mathcal{O}} P_t(\mathcal{O})}{Z_t(s)},\ \ \  Z_t(s) = \int e^{s\mathcal{O}}P_t(\mathcal{O})\, d\mathcal{O}.
\label{ptilt}
\end{equation}
The notation is reminiscent of that of the ensembles of equilibrium statistical mechanics, including that of the partition sum $Z_t(s)$, which is a moment-generating function that also acts as a normalization factor (there is in fact a strong connection between the two problems \cite{garrahan2009,Touchette2009}). Values of  $\mathcal{O}$ greater (smaller) than $\langle \mathcal{O} \rangle$ are thus favored by choosing a positive (negative) value of the tilting parameter $s$. In fact, the whole statistics of the observable $\mathcal{O}$ is encoded in the scaled cumulant generating function (SCGF) $\theta(s) = \lim_{t\to \infty} t^{-1} \log Z_t(s)$. Specifically, the derivatives of $\theta(s)$  of a given order correspond (up to a rescaling by time) to the cumulants of $\mathcal{O}$ of the same order \cite{Carollo2018,garrahan2009}.\\

The SCGF can be obtained as the largest eigenvalue of a modified Liouvillian generator, the so-called tilted generator \cite{Carollo2018}, which is a modified version of the quantum master equation in GKSL form, namely
\begin{equation}
\resizebox{\columnwidth}{!}{$\mathcal{L}_s[\cdot] = - i [H,\cdot] + \sum\limits_{j k}  \left(e^{s O_{jk}} L_{j k} \cdot L_{j k}^\dag -\frac{1}{2} \left\{L_{j k}^\dag L_{j k},\cdot \right\}\right).$}
\label{tiltliou}
\end{equation}
(In the particular case of a single jump operator for the transition $\ket{N}\to \ket{1}$, denoted $L_\textrm{link}$, with local contribution $O_{N1} = 1$, this is equivalent to Eq. (3) in the main text.)
In the sum over $j$ and $k$ we only consider those vertices that are incoherently connected by a jump operator $L_{jk}$, while the contribution to the observable $\mathcal{O}$ given by  $O_{jk}$ may be zero in all except one or a few transitions, depending on the transport problem under consideration. The generator (\ref{tiltliou}) satisfies $\mathcal{L}_s[r_s] = \theta(s)\, r_s$, where $r_s$ is the (right) eigenmatrix associated with the largest eigenvalue (SCGF). In terms of the adjoint map acting on the system operators
\begin{equation}
\resizebox{\columnwidth}{!}{$
\mathcal{L}^{\dagger}_s[\cdot] = i [H,\cdot] + \sum\limits_{j k} \left(e^{s O_{jk}} L_{jk}^\dag \cdot L_{jk} -\frac{1}{2} \left\{L_{jk}^\dag L_{jk},\cdot \right\}\right).$}
\label{tiltadjliou}
\end{equation}
the SCGF satisifes $\mathcal{L}^{\dagger}_s[l_s] = \theta(s)\, l_s$, where $l_s$ is the (left) eigenmatrix associated with the largest eigenvalue of $\mathcal{L}^\dagger_s[\cdot]$, which is also the SCGF \cite{Carollo2018}. These eigenmatrices are normalized so that $\tr[l_s r_s] = \tr[r_s] = 1$.\\

While the generator in Eq.~(\ref{tiltliou}) and its Heisenberg-picture counterpart in Eq.~(\ref{tiltadjliou}) are useful to obtain, via the SCGF $\theta(s)$, the statistics of the observable $\mathcal{O}$ for a given $s$ in Eq.~(\ref{ptilt}), they do not generate a physical dynamics. In fact, Eq.~(\ref{tiltliou}) does not preserve the trace of the density operator for $s\neq 0$ \cite{Carollo2018}. We next summarize a procedure to overcome this limitation and find a physical GKSL dynamics that naturally sustains the statistics given by Eq.~(\ref{ptilt}) in its stationary state.

\subsection*{A.3. Quantum Doob generator}

In Ref.~\cite{Carollo2018} a procedure is proposed to modify Eq.~(\ref{tiltliou}) so as to obtain a completely positive trace-preserving Markovian evolution, given by the so-called quantum Doob generator, with $P_t^s(\mathcal{O})$ in Eq.~(\ref{ptilt}) as its stationary-state distribution for the observable $\mathcal{O}$. As we mentioned above, this cannot be the tilted operator in Eq.~(\ref{tiltliou}) itself, as the dynamics that it generates is nonphysical. Instead, the quantum Doob generator for tilting $s \neq 0$ is
\begin{equation}
\mathcal{L}^D_s[\cdot] = \ell_s^{1/2} \mathcal{L}_s\left[ \ell_s^{-1/2}(\cdot) \ell_s^{-1/2}\right]  \ell_s^{1/2} -\theta(s) (\cdot).
\label{liouD}
\end{equation}
To construct this operator one needs the left eigenmatrix $l_s$ of the tilted dynamics, mentioned in the previous section, or more specifically its square root, $l_s^{1/2}$, satisfying $l_s^{1/2} l_s^{1/2} = l_s$. The quantum Doob generator \eqref{liouD} can be written in GKSL form as follows:
\begin{equation}
\resizebox{\columnwidth}{!}{$
\mathcal{L}^D_s[\cdot] = - i [H^D_s,\cdot] + \sum\limits_{j k }  \left( L^D_{j k,s} \cdot L^{D\, \dag}_{j k,s} -\frac{1}{2} \left\{L^{D\, \dag}_{j k ,s} L^D_{j k ,s},\cdot \right\}\right),$}
\label{liouD2}
\end{equation}
where the sum again runs over the indices of incoherently connected vertices. The effective Hamiltonian in Eq.~\eqref{liouD2} is given by
\begin{equation}
H^D_s = \frac{1}{2}\, l_s^{1/2}\left(H- \frac{i}{2}\sum\limits_{j k} L^{\dag}_{j k} L_{j k} \right) l_s^{-1/2} + \text{H.c.},
\label{hd}
\end{equation}
with the sum reducing to $\sum\limits_{j} R_j |j\rangle\langle j|$, with $R_j = \sum\limits_{k} \gamma_{j k}$ (again, summing over $k$ originally connected to $j$ by an incoherent link). And the effective jump operators are given by 
\begin{equation}
 L^D_{j k,s} = e^{(s/2) O_{jk}}\, l_s^{1/2}\, L_{jk}\, l_s^{-1/2}.
 \label{ld}
\end{equation}
These may now include linear combinations of several incoherent transitions, including those that were not present in the original dissipator. The latter is a feature of the quantum Doob transform, which is absent from the classical Doob transform of random walks on graphs \cite{coghi2019,gutierrez21a}. The general expressions given in Eqs.~\eqref{hd} and \eqref{ld} reduce to Eqs.~(5) and (6) in the main text for the model that is considered there.

The evolution given by Eq.~(\ref{liouD2}) can be shown to have a steady state $\rho^\text{st}_s =  l_s^{1/2} r_s\, l_s^{1/2}$, where the statistics of the observable $\mathcal{O}$ follows $P_t^s(\mathcal{O})$ as given in Eq.~(\ref{ptilt}). For the particular case of $s=0$, Eq.~(\ref{liouD2}) with the definitions given in Eqs.~(\ref{hd}) and (\ref{ld}) reduces to the original quantum master equation in GKSL form; see Eq. (1) in the main text.

\section*{Appendix B. MonteCarlo analysis}
\label{appB}

\subsection*{B.1. Generation of Hamiltonians}

In order to perform the Monte Carlo analysis, a set of Hamiltonians $\displaystyle\{\,H^{(n)}\,\}_{n=1}^{10^4}$ is generated using \texttt{Mathematica}. Each Hamiltonian is constructed as a Hermitian matrix of size $7\times 7$, following the procedure bellow:
\begin{enumerate}[left=0pt]
    \item First, a real matrix $X^{(n)}$ is created by randomly generating each entry such that $X^{(n)}_{ij}\in[1,216]$, under a uniform distribution.
    \item Then, $X^{(n)}$ is used to build a random Hermitian matrix $Y^{(n)}$ (hence symmetric, as it is real) via the relation $Y^{(n)}=\frac{1}{2}(X^{(n)}+(X^{(n)})^T)$.
    \item Finally, using the Hermitian matrix $Y^{(n)}$, and manually setting the  interaction between nodes $1$ and $N$ to be the weakest, the Hamiltonian $H^{(n)}$ is defined as: 
    \begin{equation*}
        H^{(n)}_{ij}=\left\{
        \begin{array}{lll}
            &1\,, & \mbox{if} \;\;\;(i,j)=(1,N)\;\;\mbox{or}\;\;(N,1)\\
            &Y^{(n)}_{ij}\,, & \mbox{otherwise}.
        \end{array}
        \right.
    \end{equation*}
\end{enumerate}

\subsection*{B.2. Analysis of the Doob transform}

\begin{figure*}[htbp]
    \centering
    \includegraphics[width=.36\linewidth]{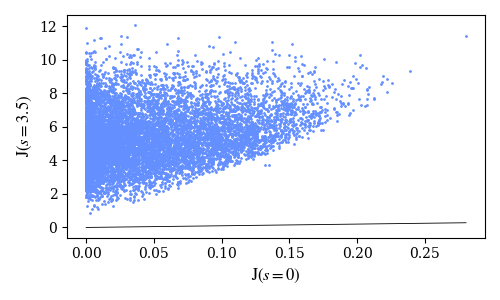}\\    
    \includegraphics[width=.36\linewidth]{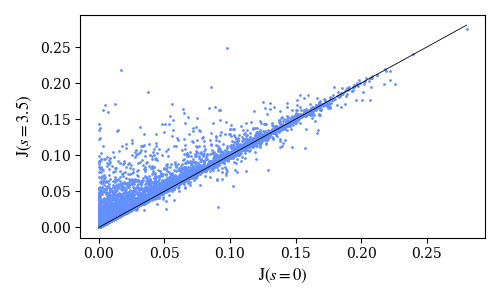}
    \includegraphics[width=.36\linewidth]{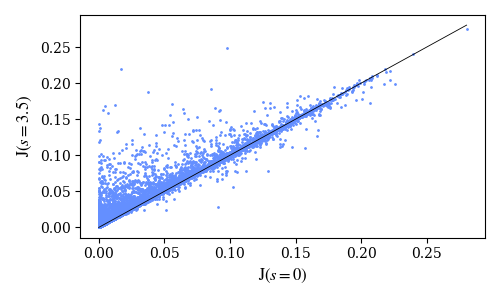}
    \caption{Top: Original (x-axis) versus Doob-transformed efficiencies with $s=3.5$ (y-axis).  Bottom Left: The same but using the Doob Hamiltonian and the original dissipation. Bottom Right: The same as the center panel but with original dissipation and Doob-Hamiltonian modified so that $(H_s^D)_{1N} = 1$. All plots are made for $10^4$ random initial Hamiltonians and the solid line has slope one, corresponding to $J({s=0})=J({s=3.5})$.}
    \label{fig:correlations}
\end{figure*}

In Fig.~\ref{fig:correlations} the efficiency of the randomly generated systems is compared with the corresponding Doob-transformed system for $s=3.5$. In the top plot we can see the full Doob-transformed efficiency, which improves for all cases and in most of them increases by more than an order of magnitude. These results can be partially explained by the fact that the dissipative part is modified so that the rate between the output and the input node is increased. In the bottom left plot analogous results are displayed based on the Doob Hamiltonian but changing the dissipative dynamics to be the original one. Finally, the bottom right plot displays the same information but with the connection between the output and the input node in the Doob Hamiltonian restored to its original value, as well as the dissipative dynamics. These three choices correspond to the different panels in Fig. 4 of the main text. One general trend is that the systems that originally are less efficient are more prone to improving than the ones with an originally high efficiency. This result aligns with previous ones concerning noise-enhanced quantum transport \cite{scholak:pre11,manzano:po13}.

We have also performed a statistical analysis of the relation between centrosymmetry and efficiency by calculating the joint probabilities of having a simultaneous increase in centrosymmetry $\varepsilon$ and average current $J(s) = \theta'(s)$, $P(\uparrow \varepsilon, \uparrow J)$, as well as the conditional probability of having a current increase conditioned on a centrosymmetry increase, $P(\uparrow J \,|\uparrow \varepsilon)$, and the other way around, $P(\uparrow \varepsilon \,|\uparrow J)$. The results are summarized in the contingency Table \ref{tab:tablas doble entrada}. In the first case, we have the full Doob-transformed system, where the current is always increased and centrosymmetry is increased in $71.46\%$ of the systems. Then, is is trivial to calculate the joint and conditional probabilities
\begin{align*}
    &\rm P(\uparrow \varepsilon , \uparrow J)= P(\uparrow \varepsilon \,|\uparrow J)=71.46\%\,,   \\ 
    & \rm P(\uparrow J \,|\uparrow \varepsilon)=100\%\,.
\end{align*}
For the second case, in which in order to calculate the current the Doob-Hamiltonian is used and the dissipator is set to be the original one, see Table \ref{tab:tablas doble entrada} (center), the joint and conditional probabilities are
\begin{align*}
    \rm P(\uparrow \varepsilon , \uparrow J)&= 60.75\%\,, \\
    \rm P(\uparrow \varepsilon \,|\uparrow J)&=70.80\%\,,   \\ 
    \rm P(\uparrow J \,|\uparrow \varepsilon)&=85.01\%\,.
\end{align*}
Lastly, when the dissipator is restored to its original form and the Doob-Hamiltonian is modified so that $(H_s^D)_{1N}=1$, see Table \ref{tab:tablas doble entrada}(right), the probabilities are
\begin{align*}
    \rm P(\uparrow \varepsilon \, \uparrow J)&= 58.39\%\,, \\
    \rm P(\uparrow \varepsilon \,|\uparrow J)&=70.81\%\,,   \\ 
    \rm P(\uparrow J \,| \uparrow \varepsilon)&=81.63\%\,.
\end{align*}
In conclusion, current and centrosymmetry are highly dependent on each other: when one of them increases, most likely the other does too in any of the aforementioned cases. 

\begin{table}[htbp]
    \footnotesize
    \centering
\begin{tabular}{c|c|c|c}
\cline{2-3}
                                    & $\uparrow \rm J$& $\downarrow \rm J$ &                         \\ \hline
\multicolumn{1}{|c|}{$\uparrow \rm \varepsilon$}  & {\scriptsize 7146}  & \scriptsize 0  & \multicolumn{1}{c|}{\scriptsize 7146}  \\ \hline
\multicolumn{1}{|c|}{$\downarrow \rm \varepsilon$} &  \scriptsize 2854 & \scriptsize 0 & \multicolumn{1}{c|}{\scriptsize 2854} \\ \hline 
                                    &  \scriptsize$10^4$              & \scriptsize 0             & \multicolumn{1}{c|}{\scriptsize $10^4$} \\\cline{2-4} 
\end{tabular}\quad
\begin{tabular}{c|c|c|c}
\cline{2-3}
                                    & $\uparrow \rm J$ & $\downarrow \rm J$ &                         \\ \hline
\multicolumn{1}{|c|}{$\uparrow \rm \varepsilon$}  & \scriptsize 6075  & \scriptsize 1071  & \multicolumn{1}{c|}{\scriptsize 7146}  \\ \hline
\multicolumn{1}{|c|}{$\downarrow \rm \varepsilon$} & \scriptsize 2506 & \scriptsize 348 & \multicolumn{1}{c|}{\scriptsize 2854} \\ \hline 
& \scriptsize 8581  & \scriptsize 1419   & \multicolumn{1}{c|}{\scriptsize $10^4$} \\\cline{2-4} 
\end{tabular}\quad
\begin{tabular}{c|c|c|c}
\cline{2-3}
                                    & $\uparrow \rm J$ & $\downarrow \rm J$ &                         \\ \hline
\multicolumn{1}{|c|}{$\uparrow \rm \varepsilon$}  & \scriptsize 5839  & \scriptsize 1314  & \multicolumn{1}{c|}{ \scriptsize7153}  \\ \hline
\multicolumn{1}{|c|}{$\downarrow \rm \varepsilon$} & \scriptsize 2407 & \scriptsize 440 & \multicolumn{1}{c|}{\scriptsize 2847} \\ \hline 
 & \scriptsize 8246   & \scriptsize 1754   & \multicolumn{1}{c|}{\scriptsize $10^4$} \\\cline{2-4} 
\end{tabular}
    \caption{Contingency tables describing the number of cases that satisfy the pairs given by increasing ($\uparrow$) or decreasing ($\downarrow$) both current ($J$) and centrosymmetry $(\varepsilon)$. The three correspond to the different ways to calculate the current. Left: Doob-transformed efficiency. Center: Doob Hamiltonian and original dissipator. Right: Doob Hamiltonian modified so that $H^D_{1N}=1$ and original dissipator.}
    \label{tab:tablas doble entrada}
\end{table}

\section*{Appendix C. Double-excitation problem}
\label{sec:appc}

\subsection*{C.1 Model}

To generalize our results, here we study the specific case of a network with two excitations. The system state in the Fock basis takes the form $\ket{ij}\in \mathcal{H}_2$, living in a Hilbert space of $\text{dim}(\mathcal{H}_2)=N(N-1)$. The index $i(j)$ represents the first(second) particle position and we impose an exclusion principle, meaning that both particles cannot occupy the same space $i\neq j$. Considering only one-particle jumps, a general Hamiltonian can be derived from the single particle one as 

\begin{equation}
H_2 = \sum_{\substack{ i,j,k=1 \\(i \neq j \neq k)}}^N 
J_{ik} \left(\op{ij}{kj} + \op{ji}{jk} \right) + \text{H.c.}
\end{equation}
where the parameters $J_{ij}$ come from the same $10^4$ random Hamiltonians as the single particle case.

As both particles are identical, this Hamiltonian is degenerate.  To avoid these degeneracies, and to reduce the dimensionality, we consider only the symmetric manifold. States are then defined as 
\begin{equation*}
    \ket{ij}_S = \frac{1}{\sqrt{2}}(\ket{ij}+\ket{ji})\quad\mbox{with}\quad i<j\,,
\end{equation*}
the dimension of the symmetrized Hilbert space for two particles being equal to $\frac{N(N-1)}{2}$.

In this manifold, the Hamiltonian reads
\begin{eqnarray}
    H_{\text{2-sym}} &=&  \sum_{\substack{ i < j  \\ k < j}} J_{ik} \op{ij}{kj}_S + \sum_{\substack{ i < j  \\ i < k }} J_{jk} \op{ij}{ik}_S \nonumber\\
    +&&\sum_{\substack{ i < j  \\  j<k }} J_{ik} \op{ij}{jk}_S + \sum_{\substack{ i < j  \\  k<i }} J_{jk} \op{ij}{ki}_S.  \nonumber\\
\end{eqnarray}
%
The Lindblad operator for symmetric states is
\begin{equation*}
    L = \sum_{i>1}^{N-1}\ket{1\,i}_S\bra{i\,N}_S\,.
\end{equation*}

For clarification, this would result in the following Hamiltonian and jump operator for the $N=4$ case:
\begin{equation*}
\resizebox{\columnwidth}{!}{$
    H=
    \begin{pmatrix}
        0 & J_{23} & J_{24} & J_{13} & J_{14} & 0 \\
        J_{23} & 0 & J_{34} & J_{12} & 0 & J_{14} \\
        J_{24} & J_{34} & 0 & 0 & J_{12} & J_{13} \\
        J_{13} & J_{12} & 0 & 0 & J_{34} & J_{24} \\
        J_{14} & 0 & J_{12} & J_{34} & 0 & J_{23} \\
        0 & J_{14} & J_{13} & J_{24} & J_{23} & 0
     \end{pmatrix}
     \phantom{m} \mbox{and}\phantom{m} L=
     \begin{pmatrix}
        0 & 0 & 1 & 0 \\
        0 & 0 & 0 & 1 \\
        0 & 0 & 0 & 0 \\
        0 & 0 & 0 & 0
     \end{pmatrix}\,.$}
\end{equation*}

\subsection*{C.2 Transport optimization}

\begin{figure}
    \includegraphics[width=1\linewidth]{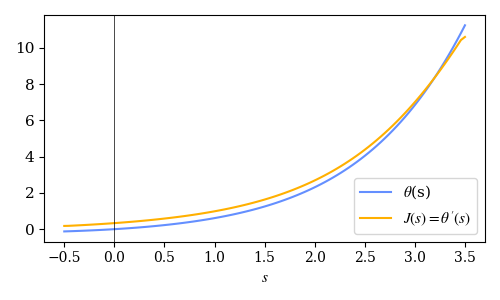}
    \caption{Scaled cumulant-generating function $\theta(s)$ and current $J(s) = \theta'(s)$, as a function of the tilting parameter $s$.}
    \label{fig:DoubleEx_Large_Deviation}    
\end{figure}

Using this framework we have analyzed the behavior of a network with $N=7$ nodes, two particles and the same coupling as in the main set. In Fig.~\ref{fig:DoubleEx_Large_Deviation}, the SCGF and the flux of excitations as a function of the counting field $s$ are displayed. The results are very similar to the ones obtained for the 1-excitation case (see Fig.~2 of main text, top). We conclude that the number of particles does not affect the phenomenology of the problem. 

\begin{figure*}
    \centering
    \includegraphics[width=.32\linewidth]{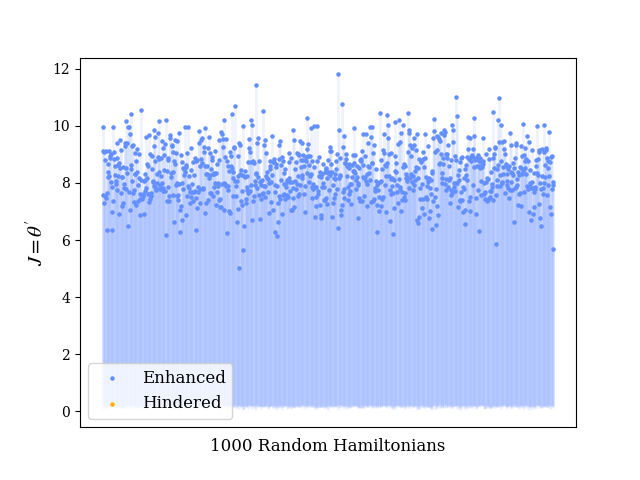}    
    \includegraphics[width=.32\linewidth]{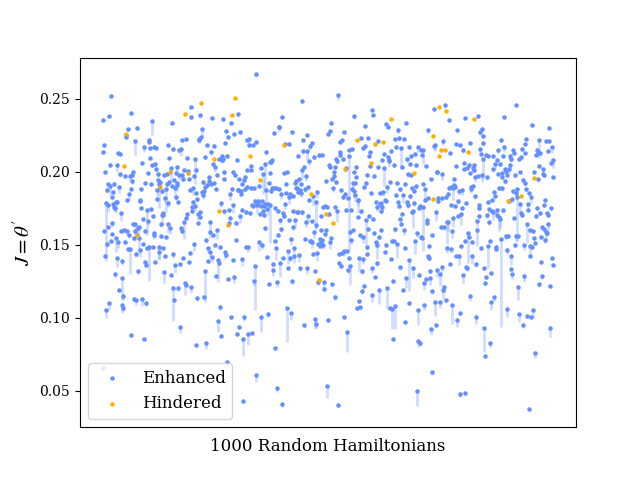}
    \includegraphics[width=.32\linewidth]{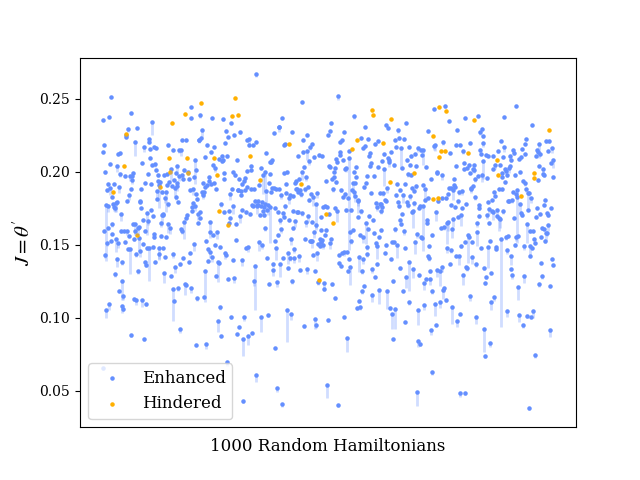}
    \caption{In all plots vertical segments start at the original efficiency and end at the Doob-transformed efficiency, highlighted by a dot, with colors indicating whether the efficiency is enhanced or hindered (see legend). Left: Original and Doob-transformed efficiency, using both the Doob Hamiltonian, Eq. (4) of the main text, and the Doob jump operator, Eq. (5) of the main text.  Center: Original and the Doob Hamiltonian together with the original jump operator $L_\textrm{link}$. Right: Original and, systems with $L_\textrm{link}$ as jump operator and the Doob Hamiltonian modified so that $(H_s^D)_{\alpha\beta} = 1$ for $(\alpha,\beta)\in\mathcal{S}$. All plots are based on 1000 random initial Hamiltonians.}
    \label{fig:DoubleEx_Rain}
\end{figure*}

Finally, results in Fig.~4 of the main text, showing the improvement of $1000$ random Hamiltonians when we apply a Doob transform with $s=3.5$  are replicated for the two particles case in Fig.~\ref{fig:DoubleEx_Rain}. Fig.~\ref{fig:DoubleEx_Rain} (Left) shows the improvement of the efficiency when both Hamiltonian and jump operators are transformed. As in the single excitation case, it shows a $100\%$ improvement as a consequence of the intrinsic properties of the quantum Doob transform. For the case when only the Hamiltonian changes but the incoherent part of the dynamics is fixed, Fig.~\ref{fig:DoubleEx_Rain} (Center), we get an improvement for $94.82\%$ of the initial configurations. Finally, considering that the Hamiltonian changes but the connection between input and output nodes remain constant (C.1), Fig.~\ref{fig:DoubleEx_Rain} (Right), we obtain an improvement for $93.63\%$ of the initial configurations. These percentages indicate an even stronger improvement than in the single-excitation case discussed in the main text.

In conclusion, the original problem is generalizable to the double-excitation situation, with consistent results. We expect that our results will remain consistent (with some quantitative differences) for an arbitrary number of excitations.

\subsection*{C.3 Setting of the constant connection between input and output}
Whereas in the one-excitation problem the constant connection between input and output was only a matter of setting $(H_s^D)_{1N} = 1$ \textit{a posteriori}, in the two-excitation problem (see Fig.~\ref{fig:DoubleEx_Rain}, Right) it is not as trivial. In this case, we need to impose
\begin{equation}
    H_{\alpha\beta} = 1\quad \Leftrightarrow \quad(\alpha,\beta)\in\mathcal{S}\,,
\label{ones}
\end{equation}
where $\mathcal{S}$ is the set of pairs of indexes that come from input-output transitions.

For a generic transition $\ket{mn}_S\;\rightarrow \;\ket{ij}_S$, one needs to identify the relation between the pairs $(m,n)$ and $(i,j)$ with the pair $(\alpha,\beta)$. It is straightforward to see that they are related by a function $f$ such that $\alpha=f(m,n)$ and $\beta=f(i,j)$. Solving a classic induction problem, we obtain that the form of $f$ is
\begin{equation*}
    f(m,n)=(m-1)N+n-\frac{(m+1)m}{2}\,.
\end{equation*}
The two cases of interest are the transitions that satisfy either $\{m=i,n=N,j=1\}$ or $\{n=j,m=N,i=1\}$. Hence, we can write $\mathcal{S} = \mathcal{S}_1\cup\mathcal{S}_2$, where
\begin{eqnarray*}
    &\mathcal{S}_1 & =\{\,(\alpha,\beta)\; / \; \;
    \begin{split}
        &\alpha = f(i,N) \\
         &\beta = f(i,1)
    \end{split} \,,\; \; 2\leq i \leq N-1\,\} \\
    \mbox{and} \\
     &\mathcal{S}_2& =\{\,(\alpha,\beta)\; / \; \;
    \begin{split}
        &\alpha = f(N,j) \\
         &\beta = f(1,j)
    \end{split} \,,\; \; 2\leq j \leq N-1\,\}\,.
\end{eqnarray*}
From these expressions it is now easy to conclude that the number of entries set to $1$ [see Eq.~(\ref{ones})] in the \textit{a posteriori} Hamiltonian depends on the size of the network as $\#\mathcal{S}= 2(N-2)$.

\bibliography{biblio.bib}

\end{document}